\documentstyle[prl,aps,epsf]{revtex}

\begin{document}
\draft
\title{Interaction Energy of `t Hooft-Polyakov Monopoles}
\author{{\large B. Kleihaus}
  \footnote{
on leave of absence from
Fachbereich Physik, Universit\"at Oldenburg, D-26111 Oldenburg, Germany
           }
}
\address{
Department of Mathematical Physics, National University of Ireland Maynooth
}
\author{{\large J. Kunz}}
\address{
Fachbereich Physik, Universit\"at Oldenburg, D-26111 Oldenburg, Germany
}
\author{and} 
\author{{\large D. H. Tchrakian}}
\address{
Department of Mathematical Physics, National University of Ireland Maynooth\\
and \\
School of Theoretical Physics -- DIAS, 10 Burlington Road, Dublin 4, Ireland
}

\date{\today}
\newcommand{\dd}{\mbox{d}}\newcommand{\tr}{\mbox{tr}}
\newcommand{\ee}{\end{equation}}
\newcommand{\be}{\begin{equation}}
\newcommand{\ii}{\mbox{i}}\newcommand{\e}{\mbox{e}}
\newcommand{\pa}{\partial}\newcommand{\Om}{\Omega}
\newcommand{\vep}{\varepsilon}
\newcommand{\bfph}{{\bf \phi}}
\newcommand{\lm}{\lambda}
\def\theequation{\arabic{equation}}
\renewcommand{\thefootnote}{\fnsymbol{footnote}}
\newcommand{\re}[1]{(\ref{#1})}
\newcommand{\bfR}{{\sf R\hspace*{-0.9ex}\rule{0.15ex}%
{1.5ex}\hspace*{0.9ex}}}
\newcommand{\N}{{\sf N\hspace*{-1.0ex}\rule{0.15ex}%
{1.3ex}\hspace*{1.0ex}}}
\newcommand{\Q}{{\sf Q\hspace*{-1.1ex}\rule{0.15ex}%
{1.5ex}\hspace*{1.1ex}}}
\newcommand{\C}{{\sf C\hspace*{-0.9ex}\rule{0.15ex}%
{1.3ex}\hspace*{0.9ex}}}
\renewcommand{\thefootnote}{\arabic{footnote}}

\maketitle
\begin{abstract}
The dependence of the energies of axially symmetric monopoles of
magnetic charges 2 and 3,
on the Higgs self-interaction coupling constant, is studied numerically.
Comparing the energy per unit topological charge of the charge-2 monopole with
the energy of the spherically symmetric charge-1 monopole, we confirm
that there is only a repulsive phase in
the interaction energy between like monopoles
\end{abstract}

\vfill
\noindent {Preprint hep-th/9804192} \hfill\break
\vfill\eject

It is well known that like monopoles \cite{tH,P} of the Georgi-Glashow (GG)
model
exhibit only a repulsive phase. In the  Bogomol'nyi-Prasad-Sommerfield 
(BPS) limit \cite{B,PS}, when the
dimensionless coupling constant $\lambda$ parametrising the strength of the
Higgs
self-interaction potential vanishes, this interaction disappears. This is a
consequence of the vanishing of all components of the stress tensor \cite{JT}
in this limit. In the BPS limit the Higgs field becomes massless and mediates a
long range attractive force which cancels the long range repulsive
magnetic force of the $U(1)$ field, exactly. The force between like monopoles
in the BPS limit was studied in detail by Manton \cite{M} and
Nahm \cite{N}, who showed
that it decreases faster than any inverse power. When $\lambda >0$ however,
the Higgs field becomes massive and as a result decays exponentially.
Consequently
the long range magnetic field dominates at large distances,
leading to the repulsion of like monopoles of the GG model.
This was concluded by Goldberg et al.~\cite{Gold}, using the time-rate 
of change of the stress tensor for the field configuration of two 
exponentially localized monopoles, situated apart at a distance much larger
than the sizes of  the monopole cores.

Inspite of the well known scenario described above, to our knowledge, there has
not been any detailed numerical verification of it to date. It is the
purpose of the present work to supply this.

A numerical study of the energy of the {\it unit} charge spherically
symmetric `t Hooft-Polyakov monopole was carried out
by Bogomol'nyi and Marinov long ago \cite{BM},
where the dependence of the energy on $\lambda$ was studied in detail. No such
study was carried out for axially symmetric monopoles of higher charges. The
study of axially symmetric monopoles was
confined to the BPS limit only \cite{RR,FHP}.
In the present work we will analyse the $\lambda >0$ case, and in particular for
the $n=2$ and $n=3$ monopoles of the GG model.

Our procedure is analogous to the work of Jacobs
and Rebbi \cite{JR}, where they study the energy {\it per unit charge}
of the topologically stable charge-$n$ vortices of the Abelian Higgs model
(AHM) \cite{JT},
and demonstrate that in these theories there are bound states of vortices
or more
generally, that there are both attractive and repulsive phases.

In both the
AHM and the GG model at hand, the masses of the solitons, the
vortex (resp.) the monopole, depend on the strength of the dimensionless Higgs
self-interaction coupling constant $\lambda$. In both cases,
topological stability is
guaranteed if $\lambda >0$. The difference between these two systems is that
the values of $\lambda$ for which the topological inequalities are saturated,
namely the critical values $\lambda_c$, are different.
For the AHM $\lambda_c =1$ while for the GG model $\lambda_c =0$. It is
clear that
in the BPS limit ($\lambda =\lambda_c $) the masses {\it per unit charge} of
charge-$n$ solitons in either of these two theories are equal to $1$,
for all $n$.

It follows that in the case of the AHM the
energy {\it per unit charge} vs. $\lambda$ curves for the $n=1$ and $n=2$
vortices cross at least at one point, namely at $\lambda_c =1$, and hence that
there will be both attractive/repulsive phases according as to whether
the mass {\it per unit charge} of the $n=2$ vortex is lower/higher than that of
the $n=1$ vortex. That there is only one such intersection point can be
speculated
on the grounds that we expect such curves for solitons to be monotonic. This
scenario was confirmed by the numerical analysis of Ref.~\cite{JR}.

In the case of the GG model, the $n=1$ and $n=2$ curves meet at
$\lambda_c =0$. Speculating again that that there is only one such intersection,
we would conclude that there is only an attractive/repulsive phase
according as to whether
the mass {\it per unit charge} of the $n=2$ monopole is lower/higher than
that of
the $n=1$ monopole. Speculating further, on the basis of the results found in
Refs.~\cite{M,N}, we would expect that there is only a repulsive
phase in the GG models. This expected result is precisely what we will verify
by numerical analysis in the present work.

In addition to confirming this expected conclusion,
we will find the detailed dependences of the monopole masses {\it per unit
charge} of the axially symmetric
charge-$n$ monopole on the Higgs coupling constant $\lambda$.
The numerical analysis will be carried out for $n=2$ and $n=3$. In this sense,
the present work is a charge-$n$ version of the work of Ref.~\cite{BM}.

The static Hamiltonian of the GG model is 
\begin{equation}
{\cal H} = 
\frac{1}{2} Tr( |F_{ij}|^2 )
+Tr( |D_i \phi|^2 )
+\frac{1}{16} \lambda ( 2 Tr(\phi^2) + \eta^2)^2 \ ,
\end{equation}
with the field strength tensor 
\begin{equation}
F_{ij} = \partial_i A_j - \partial_j A_i + [A_i, A_j]
\end{equation}
of the gauge field $A_i $,
and the covariant derivative
\begin{equation}
D_i \phi = \partial_i \phi +  [A_i, \phi ] \ ,
\end{equation}
of the Higgs field $ \phi$, and
$\eta$ denotes the vacuum expection value of the Higgs field.
The mass of the vector fields is $M_W=\eta$ (we have set the gauge coupling constant
to one),
and the Higgs coupling constant $\lambda$ is related to the Higgs mass $M_H$ by
$\lambda= 2 M^2_H/M^2_W$.

Employing spherical coordinates, we parametrize
the non-vanishing components of the gauge field by the Ansatz 
\[
A_r = \frac{H_1(r,\theta)}{r} \ \frac{i \tau_{\varphi}^{(n)}}{2}\ , \ \ 
A_\theta = ( 1 - H_2(r,\theta)) \ \frac{i \tau_{\varphi}^{(n)}}{2}\ , \ \
\nonumber
\]
\begin{equation}
A_\varphi = -n \sin \theta \ [ H_3(r,\theta) \ \frac{i \tau_{r}^{(n)}}{2}
               +(1 - H_4(r,\theta)) \ \frac{i \tau_{\theta}^{(n)}}{2}] \ .              
\label{Aphi}
\end{equation}
and the Higgs field by
\begin{equation}
\phi = \phi_1(r,\theta) \ \frac{i \tau_{r}^{(n)}}{2} 
     + \phi_2(r,\theta) \ \frac{i \tau_{\theta}^{(n)}}{2} \ .
\end{equation}
The matrices $\tau_{i}^{(n)}$ are defined by
$$
\tau_{r}^{(n)}      = \sin \theta \cos n\varphi \ \sigma_1 
                     +\sin \theta \sin n\varphi \ \sigma_2
                     +\cos \theta \ \sigma_3 \ ,
$$
$$
\tau_{\theta}^{(n)} = \cos \theta \cos n\varphi \ \sigma_1 
                     +\cos \theta \sin n\varphi \ \sigma_2
                     -\sin \theta \ \sigma_3   \ ,               
$$
$$
\tau_{\varphi}^{(n)}   =  -\sin n\varphi \ \sigma_1 
                       +\cos n\varphi \ \sigma_2 \ ,
$$
where $\sigma_i$ denote the Pauli matrices and the 
integer $n$ refers to the winding number which equals
the topological charge of the solutions.
This Ansatz (based on the Ansatz used in Ref.~\cite{RR})
allows for axially symmetric solutions with $n>1$ and reduces to the 
Ansatz for spherically symmetric solutions if $n=1$. 
Due to a local $U(1)$ invariance of the Ansatz it is 
necessary to impose a gauge fixing condition which guarantees the uniqueness of 
solutions. Here we take the gauge condition
$ r \partial_r H_1 - \partial_\theta H_2 = 0$ \cite{KKB}.

The energy of a solution with winding number $n$
\begin{equation}
E_n(\lambda) = \int {\cal H} d^3r  \ge 4 \pi Q \eta
\end{equation}
is bounded from below by the magnetic flux
\begin{equation}
 \varepsilon_{ijk} \int Tr( F_{ij}D_k \phi) d^3r 
= \varepsilon_{ijk} \int Tr(\phi F_{ij} ) dS_k  = 4 \pi Q \eta \ ,
\end{equation}
where $Q$ represents the topological charge,
and $Q=n$ for the Ansatz given above, provided that the solutions
satisfy the appropriate boundary conditions (stated in the following).

The boundary conditions follow from the requirements
of finite energy and analyticity as well as symmetry.
They are given by

$$H_1(0,\theta)=H_3(0,\theta)=\phi_1(0,\theta)=\phi_2(0,\theta)=0
\ , \ \ H_2(0,\theta)=H_4(0,\theta)=1$$
at the origin,
$$H_i(\infty,\theta)=\phi_2(\infty,\theta)=0
\ , \ \
        \phi_1(\infty,\theta)/\eta=1 $$
at infinity,
$$H_1(r,0)=H_3(r,0)=\phi_2(r,0)=0
\ , \ \
                 \partial_\theta H_2(r,0)=\partial_\theta H_4(r,0)=
                 \partial_\theta\phi_1(r,0)=0 $$
on the $z$-axis and 
$$H_1(r,\pi/2)=H_3(r,\pi/2)=\phi_2(r,\pi/2)=0
\ , \ \
                 \partial_\theta H_2(r,\pi/2)=\partial_\theta H_4(r,\pi/2)=
                 \partial_\theta\phi_1(r,\pi/2)=0 $$
on the $\rho$-axis. \vspace{1.cm} \\

\begin{figure}
\centering
\vspace{-1cm}
\mbox{  \epsfysize=7cm \epsffile{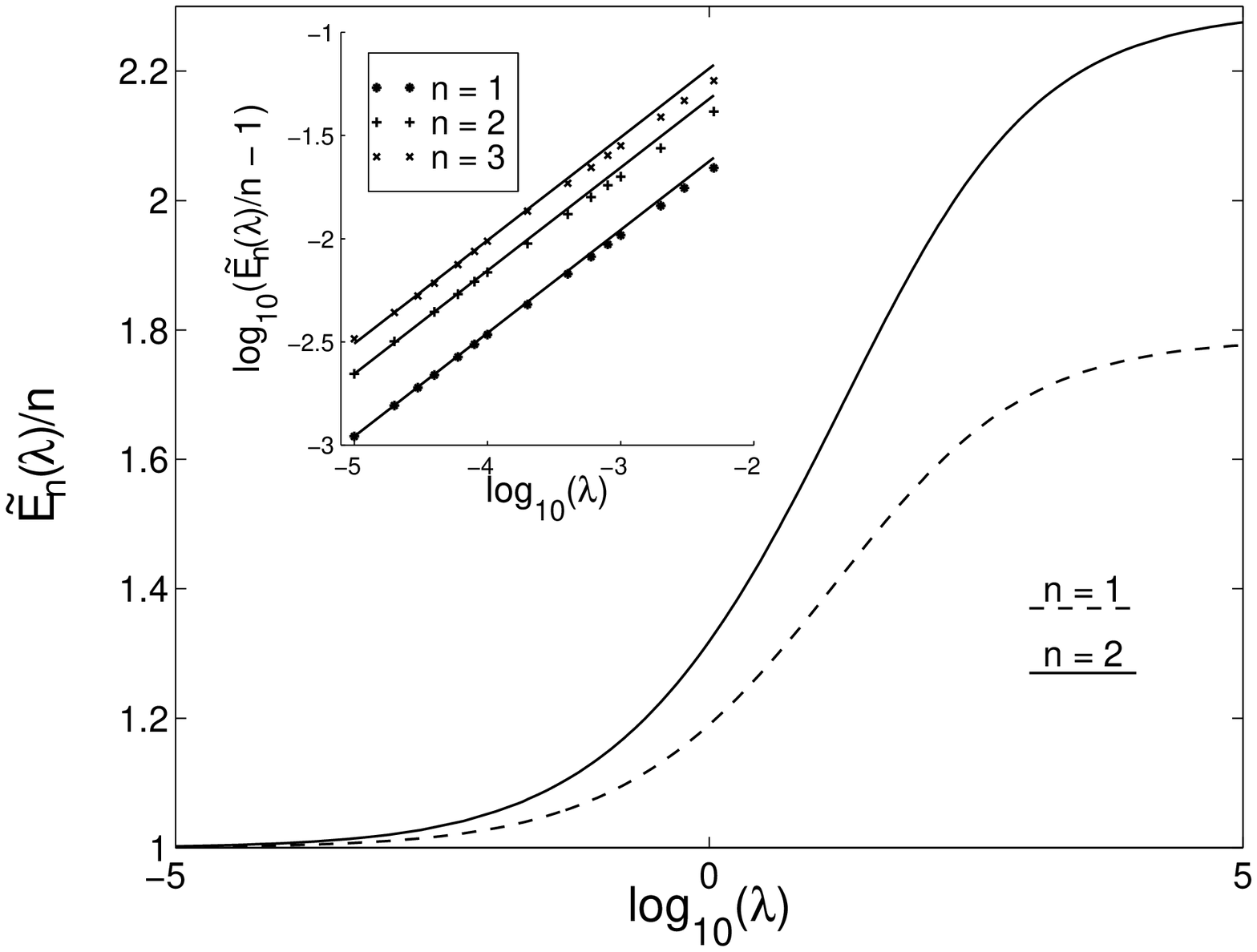}}\\
\noindent Fig.~1\\
The normalized energy {\it per unit charge} as a function of $\lambda$ is shown for 
$n=1$ (dashed line) and $n=2$ (solid line). 
The inlet shows the excess of the energy {\it per unit charge}
for small values of $\lambda$ for $n=1$ ($\ast$), $n=2$ ($+$) and $n=3$ ($\times$).
The straight lines with slope $1/2$ are fitted to the data values 
corresponding to $\lambda = 1 \times 10^{-5}$ for $n=1$ and $n=2$ and to
$\lambda = 2 \times 10^{-5}$ for $n=3$.
\end{figure}

The partial differential equations for the functions 
$H_i(r,\theta)$ and $\phi_i(r,\theta)$
are obtained by inserting the Ansatz into the Euler-Lagrange equations 
of the static Hamiltonian with gauge fixing term added.
We have solved the partial differential equations 
numerically, subject to the above
boundary conditions for the winding numbers $n=1-3$ and for 
a large range of values of the Higgs coupling constant $\lambda$. 

In Fig.~1 we show the normalized energy {\it per unit charge}
$ \tilde{E}_n(\lambda)/n = E_n(\lambda)/(4\pi n \eta)$
for $n=1$ and $n=2$ as a function of $\lambda \in [10^{-5},10^{5}]$.
In the BPS limit $\lambda=0$, both curves coincide
since $\tilde{E}_n(0)/n =1$ for all $n$ (self-dual solutions).
For $\lambda > 0 $ both curves increase monotonically,
with $ \tilde{E}_2(\lambda)/2 > \tilde{E}_1(\lambda) $
in agreement with the results of Ref.~\cite{Gold}.
In the limit of infinitely large Higgs coupling constant
the curves for $n=1$ and $n=2$ converge to the constants 
$\tilde{E}_1(\infty)=1.787$ \cite{BM,BFM} 
and $ \tilde{E}_2(\infty)/2=2.293$, respectively.
Thus the two curves coincide only at $\lambda=0$,
implying the existence of only a repulsive phase.

In the limit $\lambda \rightarrow \infty$
the Higgs potential acts like a constraint, forcing
the norm of the Higgs field
\[
 |\phi(\vec{r})| = \sqrt{-2Tr(\phi^2(\vec{r}))}
 = \sqrt{\phi_1^2(r,\theta)+\phi_2^2(r,\theta)}
\]
to be equal to its vacuum expectation value,
$|\phi(\vec{r})| \equiv \eta$.
However, this is in conflict with the boundary conditions at the origin.
Thus, in this limit the norm of the
Higgs field approaches
$\eta$ for all $r$ except for $r=0$, where it vanishes.

For $n=1$ the function $\phi_2$ vanishes and
the Higgs field is parametrised by a single function which 
corresponds to its norm. 
In the limit $\lambda \rightarrow \infty$,
in this case the Higgs field itself is completely fixed 
to its vacuum expectation value for all $r$ except for $r=0$.
For $n>1$ on the other hand, 
$\phi_2(r,\theta)$ is non-trivial and only the norm of the Higgs field
is fixed to a constant for all $r$ except for $r=0$ 
in the limit  $\lambda \rightarrow \infty$. 
Thus the Higgs field itself remains non-trivial in this limit. It can
be parametrised by the phase function 
\[
\Phi(r,\theta) = \arctan(\phi_2(r,\theta)/\phi_1(r,\theta)) \ .
\]
As in the $n=1$ case, the contributions of the Higgs potential to the 
energy are negligible in this limit, and consequently the energy
becomes independent of the Higgs coupling constant.

The gauge field functions do not develop a discontinuity in the limit of
infinitely large Higgs coupling constant. Indeed, in this limit they can be 
identified with solutions of the same model with modified boundary conditions
for the Higgs field at the origin, which allow for Higgs fields with constant
norm. We have constructed these solutions 
with $|\phi(\vec{r})| \equiv \eta$ numerically for $n=1$ and $n=2$.
Denoting these solutions by $\hat{H_i}, \hat{\phi_i}$, etc.,
their gauge field functions $\hat{H_i}$ correspond to the 
limiting functions of the gauge fields of the multimonopoles,
and, for $n>1$, their phase function $\hat{\Phi}$ corresponds to the
limiting function of the phase function $\Phi$
for all $r$ except for $r=0$.
Their energy $\hat{E_n}$ is finite and
coincides with the energy $E_n(\lambda \rightarrow \infty)$ of the 
multimonopoles.
At the origin, however, the energy density diverges and the 
gauge field functions are not analytic\footnote{This is also true for $n=1$}.
Also for $n>1$, the phase function is multivalued at the origin, i.~e. 
$\hat{\Phi}(r=0,\theta)$ is a non-trivial function of $\theta$.
Expanding the differential equation for the phase function at $r=0$
we find 
$$\hat{\Phi}(r=0,\theta) = 2 \arctan( \tan^n(\theta/2)) -\theta \ .$$
 From this unacceptable behaviour at the origin, we conclude
that these solutions are unphysical.

Let us next consider the monopoles for small values 
of the Higgs coupling constant. 
Near $\lambda=0$ the normalized energy {\it per unit charge} can be written in 
the form
\[
\tilde{E}_n(\lambda)/n \sim 1 + c_n \lambda^{\alpha} \ .
\]
We show $\tilde{E}_n(\lambda)/n -1$ for $n \le 3$
and small values of $\lambda$ in Fig.~1 (inlet).
The figure suggests that the exponent $\alpha$ does not depend on the 
winding number $n$, and 
that it has the value $1/2$ and not $1$ as might have been
expected naively from the static Hamiltonian.
This result is very natural, however, since
$(\lambda/2)^{1/2}$ is just the mass of the Higgs field
(in dimensionless units).
The physical interpretation is therefore, that the increase of the 
energy with increasing Higgs coupling constant $\lambda$ 
(for small $\lambda$)
is proportional to the mass of the Higgs field. 
We remark, that the increase of the energy for small values of $\lambda$ 
cannot be obtained by substituting the Prasad-Sommerfield solution into
the energy functional. In that case, due to the power law decay ($ \sim 1/r $) 
of the Higgs field the integration of the Higgs potential would diverge.
By considering a Taylor expansion of $\tilde{E}_n(\lambda)/n-1$ 
in terms of $\lambda$
at $\lambda=0$ and taking into account that the 
gauge fields and Higgs field depend on $\lambda$ implicitly, we find that
the first term is again the divergent integral of the Higgs potential 
evaluated with the Prasad-Sommerfield solution.
This implies, that
even for arbitrarily small values of $\lambda$ the model with finite
$\lambda$ cannot
be treated as a perturbation of the model with $\lambda=0$.

Let us now compare with the Weinberg-Salam model
of the electroweak interactions, where the Higgs field is in the
fundamental representation of SU(2).
The non-perturbative solutions of this model
are saddlepoints of the energy functional
(sphalerons and multisphalerons).
Here sphalerons represent the top of the
energy barrier between adjacent vacua \cite{KM,KKB},
and, like multimonopoles, multisphalerons are also characterised by 
their winding number $n$. 
Numerical analysis shows, that
for small values of the Higgs coupling constant,
the energy {\it per winding number} of multisphalerons ($n>1$)
is smaller than the energy of sphalerons ($n=1$),
whereas for large values of the Higgs coupling constant it is larger,
with equality occurring roughly at $M_H \approx M_W$ \cite{KK2}.
Because of their instability, however, one cannot speak 
of attractive and repulsive phases of sphalerons.

Finally we remark that, for $n>2$, multimonopoles need not be
axially symmetric. Indeed, in the BPS limit also multimonopole solutions
with only discrete symmetries 
have been found for the higher topological charges \cite{SUT}.
It remains an open challenge to construct such solutions
also for finite Higgs coupling constant.
In particular, a numerical analysis would reveal, which of the
possible shapes of a multimonopole with a given higher topological
charge would be energetically favoured.

{\bf Acknowledgements} This work was carried out under 
Basic Science Research project SC/97/636 of
FORBAIRT.

\small{

 }
\end{document}